\documentclass[twocolumn,reprint,amsmath,amssymb,aps,showpacs,eqsecnum,superscriptaddress,prx]{revtex4-1}

\usepackage{graphicx}
\usepackage{bm}
\usepackage{epsfig}
\usepackage{amssymb}
\usepackage{amsfonts}
\usepackage{braket}
\usepackage{color}
\usepackage{epstopdf}
\usepackage{hyperref}
\usepackage{float}
\restylefloat{table}
\usepackage{bibentry}
\usepackage{color}
\usepackage{multirow}
\usepackage[caption=false]{subfig}
\usepackage{ulem}

\newcommand{\magenta}[1]{{\color{magenta}{#1}}}
\newcommand{\bpm}{\begin{pmatrix}}
\newcommand{\epm}{\end{pmatrix}}
\newcommand{\ba}{\begin{eqnarray}}
\newcommand{\ea}{\end{eqnarray}}
\newcommand{\bd}{\begin{displaymath}}
\renewcommand{\v}[1]{{\bf #1}}
\newcommand{\nn}{\nonumber \\}

\graphicspath{{figures/}}

\begin{document}
\title{Consideration of Thermal Hall Effect in Undoped Cuprates}

\author{Jung Hoon Han}
\email[Electronic address:$~~$]{hanjemme@gmail.com}
\affiliation{Department of Physics, Sungkyunkwan University, Suwon 16419, Korea}
\author{Jin-Hong Park}
\affiliation{Department of Physics, Sungkyunkwan University, Suwon 16419, Korea}
\author{Patrick A. Lee}
\email[Electronic address:$~~$]{palee@mit.edu}
\affiliation{Department of Physics, Massachusetts Institute of Technology, Cambridge, Massachusetts 02139, USA }
\date{\today}

\begin{abstract} A recent observation of thermal Hall effect of magnetic origin in underdoped cuprates calls for critical re-examination of low-energy magnetic dynamics in undoped antiferromagnetic compound on square lattice, where traditional, renormalized spin-wave theory was believed to work well. Using Holstein-Primakoff boson formalism, we find that magnon-based theories can lead to finite Berry curvature in the magnon band once the Dzyaloshinskii-Moriya spin interaction is taken into account explicitly, but fail to produce non-zero thermal Hall conductivity. Assuming accidental doping by impurities and magnon scattering off of such impurity sites fails to predict skew scattering at the level of Born approximation. Local formation of skyrmion defects is also found incapable of generating magnon thermal Hall effect. Turning to spinon-based scenario, we write down a simple model by adding spin-dependent diagonal hopping to the well-known $\pi$-flux model of spinons. The resulting two-band model has Chern number in the band structure, and generates thermal Hall conductivity whose magnetic field and temperature dependences mimic closely the observed thermal Hall signals. In disclaimer, there is no firm microscopic basis of this model and we do not claim to have found an explanation of the data, but given the unexpected nature of the experimental observation, it is hoped this work could serve as a first step towards reaching some level of understanding.
\end{abstract}
\maketitle

\section{Introduction}

Traditional views on Hall effect have undergone dramatic changes over the past several decades, most prominently thanks to the observation of quantized Hall effect in two-dimensional electronic systems and subsequent realization that it is the band topology, rather than the magnetic field itself, that determines the Hall response of electronic systems~\cite{thouless82,haldane88}. It became manifest over the years, both theoretically and experimentally, that even non-electronic systems support Hall-like transport of their elementary excitations such as  photons~\cite{photon-Hall}, phonons~\cite{phonon-Hall,phonon-Hall17}, magnons~\cite{KNL,tokura10,tokura12,murakami11,zang13,lin14,skyrmion-ratchet,han-review,murakami-review}, and triplons~\cite{triplon-Hall} due to the topological character in their respective band structures or the emergent magnetic field governing their dynamics. More recently, there is growing experimental evidence of Hall-like heat (thermal) transport in magnetic materials that remain in paramagnetic, spin-liquid-like phases~\cite{ong1,ong2,matsuda16,yamashita18,matsuda18}. The physical picture regarding the origin of Hall-like phenomena for such correlated paramagnetic insulators remains poorly understood, as the Berry curvature effect only pertains to the band picture of weakly interacting quasiparticles. Schwinger-boson mean-field approximation was introduced in Refs. \cite{LHL,yamashita18} as a way to partly address the Hall effect in the paramagnetic phase. Magnetic materials exhibiting the thermal Hall effect are typically frustrated, with the pyrochlore or the kagome lattice structure~\cite{tokura10,tokura12,ong1,ong2,matsuda16,yamashita18,matsuda18} responsible for the geometric frustration, or possess significant amount of Kitaev-type interaction leading to the emergence of novel Majorana excitation~\cite{matsuda18}.

With this background, the recent observation of significant thermal Hall signal in the family of cuprate compounds comes as a surprise~\cite{taillefer}. A few salient features of the experiment may be summed up. First, the undoped antiferromagnetically ordered compound La$_2$CuO$_4$ exhibits large thermal Hall conductivity $\kappa_{xy}$ in the absence of electronic charge carriers. Phonon-related origin of $\kappa_{xy}$ is ruled out, on the ground that the spin-phonon scattering seems too weak to account for the large $\kappa_{xx}$ value in  the cuprates, and that the weak (strong) magnetic field dependence of the longitudinal (transverse) thermal conductivity $\kappa_{xx}$ ($\kappa_{xy}$) seems at odd with the phonon scenario.  Furthermore, $\kappa_{xy}$ is reduced in magnitude as doping increases, and even undergoes a sign change at some finite temperature, reflecting a mixed contribution of electronic and magnetic origins upon doping. For underdoped (and presumably undoped) La$_{2-x}$Sr$_x$CuO$_4$, the Hall effect is almost linear in the applied magnetic field $B$. Magnons, on the other hand, must have an energy gap increasing with $B$ and lead to the suppressed Hall effect at larger $B$ field. A general picture thus emerging is that the underdoped antiferromagnetic compound might have some non-trivial magnetic correlations, which are presumably gapless and revealed by the applied magnetic field through the transverse heat conduction.

What are the quasiparticles responsible for the observed transverse heat conductivity? First of all, the magnon in the experimental system has a sizable gap~\cite{taillefer}. Second, even assuming this gap to be small, we expect the gap to grow with magnetic field, whereas the thermal Hall effect initially increases with applied field. There are other objections arising from purely theoretical consideration, such as the ``no-go" theorem~\cite{KNL}, disfavoring the formation of topological Hall effect in un-frustrated square-lattice magnets. A way round this ``theorem" was invented recently~\cite{hotta19}, by adopting a model complicated enough to break spatial symmetries of the square lattice; such models do not seem to apply readily to cuprates, though. Despite these objections, we categorically look into the magnon-based scenario and add various tweaks to it, with the hope that one such model might capture the thermal Hall phenomenology. In conclusion, as we report in Secs. 2 and 3, the answer is negative; hardly any magnon-based scenario is likely to account for the thermal Hall effect in the square-lattice antiferromagnet. In Sec. 4 we outline a completely different scenario based on the spinon picture of magnetic excitation. Treating spin excitations in terms of fractionalized fermions known as spinons is an old idea, dating back to Anderson's RVB (Resonating Valence Bond) proposal. The task of applying the spinon idea to work out magnetic excitations in the cuprates was taken up in the past, notably in Refs.~\onlinecite{hsu90,ho01}. We show that a small modification of this spinon model, built around the so-called $\pi$-flux phase and its Dirac-like dispersion, can lead to finite thermal Hall conductivity with temperature and magnetic field dependences similar to the those observed~\cite{taillefer}. We emphasize that the goal of our exercise is to find a model which is capable of producing thermal Hall conductivity of the size seen by experiment. One important requirement of such model would be that the effect is linear in the applied magnetic field, as seen in the data~\cite{taillefer}; this is a feature quite naturally embodied in our model. Nevertheless, we do not claim to understand how this model can describe the cuprates. In particular we do not know how it can co-exist with N\'{e}el ordering in the insulator. We feel, however, that the experimental results are so unexpected that our modest goal can hopefully be the first step towards an explanation.

Inspired by the same experiment, a recent preprint ~\cite{sachdev18} also discussed the thermal conductivity in a spinon model, but they chose bosonic spinon and as such their treatment is complementary to our fermionic spinon model. A number of their models explicitly breaks time reversal symmetry and has net spin chirality spontaneously generated. These model will not have thermal Hall effect that is linear in magnetic field and generally speaking hysteresis may be expected.

\section{Magnon theory of thermal Hall effect in square-lattice antiferromagnet}

We begin by (re-)visiting the well-known microscopic $S=1/2$ spin Hamiltonian of the cuprates
\ba H = J \sum_{\langle ij\rangle} \v S_i \cdot \v S_{j} + \sum_{\langle ij\rangle} \v D_{ij} \cdot \v S_i \times \v S_j    - \v B \cdot \sum_i \v S_i . \nn \label{eq:HDM} \ea
In addition to the familiar spin exchange $J$, we allow the Dzyaloshinskii-Moriya (DM) interaction, originating from the small buckling of the oxygen atom out of the CuO$_2$ plane~\cite{thio88,cheong89}, and the Zeeman interaction. Sites on the square lattice are denoted simply by $i, j$, with $\langle ij\rangle$ indicating the nearest-neighbor pair of sites. The DM vectors as dictated by symmetry consideration were first worked out by Coffey {\it et al}.~\cite{coffey90}: 
\ba \v D_{i, i+\hat{x}} &=& \sqrt{2} D (-1)^i (\cos \theta_d , \sin \theta_d ), \nn
\v D_{i, i+\hat{y}} &=& -\sqrt{2} D (-1)^i (\sin \theta_d , \cos \theta_d , 0). \ea
The factor $(-1)^i = (-1)^{i_x + i_y}$ keeps track of the staggering of the DM vector. The ordered spins are forced to lie in the CuO$_2$ plane due to the DM interaction, with a small out-of-plane ferromagnetic component also dictated by the same interaction. The mean-field ansatz can be chosen as
\ba \langle \v S_i \rangle =\v n_i = n_0 \hat{z} - n_1 (-1)^i \hat{a}  . \label{eq:MF-state} \ea
It proves convenient to work with a new pair of orthonormal axes $\hat{a} = (1,1)/\sqrt{2},  \hat{b} = (-1,1)/\sqrt{2}$ instead of $\hat{x}, \hat{y}$ axes which extend along Cu-Cu directions. An orthonormal triad is formed by $\hat{a}\times\hat{b} =\hat{z}$. The mean-field energy comes out as
\ba E = 2J (n_0^2 - n_1^2) - 4 D  (\cos \theta_d - \sin \theta_d ) n_0 n_1 - B n_0 .  \nn \ea
The Zeeman energy scale at $B=10$T is only a meV, whereas the DM energy may be several meV in the cuprates. As a result, the canting angle $\theta_c$ defined as $(n_0, n_1 ) = (\sin\theta_c, \cos \theta_c)$
is dictated by the ratio $D/J$, and not so much by the Zeeman field. Minimizing the energy $E$ with respect to the canting angle $\theta_c$ gives
\ba \tan 2\theta_c = (D/J) (\cos \theta_d - \sin \theta_d ) \label{eq:canting-angle} \ea
at $B=0$. The sign of the DM energy $D$ and the angle $\theta_d$ are chosen in such a way that the canting angle is positive, $\theta_c > 0$.

Next we introduce a general formalism that allows one to convert the spin Hamiltonian (\ref{eq:HDM}) to a magnon Hamiltonian, defined around a mean-field ground state given in (\ref{eq:MF-state}). In doing so, we aim to see if the magnon theory or some of its variant can account for the thermal Hall phenomena in the undoped square-lattice antiferromagnet. The method is based on parameterizing the spin operator $\v S_i$ as
\ba \v S_i = a_i \v n_i + \v t_i , \label{eq:5.1} \ea
where $a_i$ refers to the amplitude reduction along the direction of the classical ground state spin $\v n_i$, due to the transverse fluctuation $\v t_i$. The well-known Holstein-Primakoff (HP) substitution follows from the formula
\ba a_i = S - b^\dag_i b_i , ~~  \v t_i =  t_i^\theta \bm \theta_i + t_i^\phi \bm \phi_i  , \ea
where
\ba t_i^\theta = \sqrt{S\over 2} ( b^\dag_i  + b_i ) , ~~ t_i^\phi =  i \sqrt{S\over 2}  ( b^\dag_i - b_i ) \ea
and $\bm \theta_i$ and $\bm \phi_i$ are a pair of orthonormal vectors forming the local triad $\bm \theta_i \times \bm \phi_i = \v n_i$. For this choice of triad we are guaranteed the transversality condition $\v t_i \cdot \v n_i = 0$.

Substituting (\ref{eq:5.1}) and the rest of the HP formulas into the spin Hamiltonian gives the magnon Hamiltonian,
\begin{widetext}

\ba H &=&  \sum_{\langle ij \rangle} \left[ J \bm \theta_i \cdot \bm \theta_j    + \v D_{ij} \!\cdot \!\bm \theta_i \!\times\! \bm \theta_j  \right] t_i^\theta t_j^\theta    + \sum_{\langle ij\rangle } \left[ J \bm \phi_i \cdot \bm \phi_j + \v D_{ij} \!\cdot\! \bm \phi_i \!\times\! \bm \phi_j   \right]  t_i^\phi t_j^\phi \nn
& + & \sum_{\langle ij\rangle}  \left[ J \bm \theta_i \cdot \bm \phi_j +  \v D_{ij} \!\cdot\! \bm \theta_i \!\times\! \bm \phi_j  \right] t_i^\theta t_j^\phi + \sum_{\langle ij\rangle} \left[ J \bm \phi_i \cdot \bm \theta_j  +  \v D_{ij} \!\cdot\! \bm \phi_i \!\times\! \bm \theta_j \right] t_i^\phi t_j^\theta  - \sum_i \mu_i b^\dag_i b_i , \label{eq:4.5} \ea
\end{widetext}
where
$\mu_i \!=\!  J \sum_{j\in i} \v n_i \cdot \v n_j  \!+\! \sum_{j \in i} \v D_{ij} \cdot \v n_i \times \v n_j \!-\! \v B \cdot \v n_i $.
The spin size $S=1/2$ can be absorbed by various re-definitions of the physical constants and will not be shown from now.
Our notation is such that $\langle ij\rangle$ refers to the nearest-neighbor (NN) bond, and $j\in i$ refers to the summation over the (four) NN sites $j$ that surround the site $i$. The mean-field spin configuration was already laid out in (\ref{eq:MF-state}), and we need to complete the orthonormal triad as
\ba \v n_i &=& - n_1 (-1)^i \hat{a} + n_0 \hat{z} ,  \nn
\bm \theta_i & = & n_1 (-1)^i \hat{z} + n_0 \hat{a} , \nn
\bm \phi_i & = & \hat{b} .  \label{eq:triad}\ea
This choice of parametrizing the triad is convenient because several terms in the Hamiltonian (\ref{eq:4.5}) vanish automatically: $\bm \theta_i \times \bm \theta_j = \bm 0$, $\bm \theta_i \cdot \bm \phi_j = \bm \phi_i \cdot \bm \theta_j = 0$. Remaining terms are $\bm \theta_i \cdot \bm \theta_j = -1$, $\bm \phi_i \cdot \bm \phi_j = \cos 2 \theta_c = - \v n_i \cdot \v n_j$, $\v D_{ij} \cdot \bm \phi_i \times \bm \phi_j = J \sin 2\theta_c \tan 2\theta_c $ for both $j=i+\hat{x}$ and $j=i+\hat{y}$, $\v D_{ij} \cdot \bm \theta_i \times \bm \phi_j = \v D_{ij} \cdot \bm \phi_i \times \bm \theta_j = \pm n_1 D (\cos \theta_d + \sin\theta_d )$. The $\pm$ signs refer to $j=i+\hat{x}$ and $j=i+\hat{y}$, respectively. The magnon Hamiltonian in real space becomes
\ba & H = (4J' S + Bn_0 ) \sum_i b^\dag_i b_i - J \sum_{\langle ij\rangle} t_i^\theta t_j^\theta + J' \sum_{\langle ij\rangle} t_i^\phi t_j^\phi \nn
& + D' \sum_{i} (t_i^\theta t_{i\!+\!\hat{x}}^\phi + t_i^\phi t_{i\!+\!\hat{x}}^\theta ) - D' \sum_{i} (t_i^\theta t_{i\!+\!\hat{y}}^\phi + t_i^\phi t_{i\!+\!\hat{y}}^\theta ), \nn \ea
where $J' = J/ \cos 2\theta_c$ and  $D' = n_1 D (\cos\theta_d + \sin \theta_d )$. The magnon Hamiltonian in momentum space is
\ba H & = & {1\over 2} \sum_{k} \psi^\dag_k H^0_k \psi_k  ,  \label{eq:2.27} \ea
where
\ba & \psi_k = \bpm b_{k} \\ b^\dag_{-k} \epm , ~~ H^0_k  = \bpm A_{k}  & B_k \\ B^*_k & A_{k}\epm \nn
& A_k = 4J'  + Bn_0  + (J' - J ) (\cos k_x + \cos k_y ) \nn
& B_k = - (J'+J) (\cos k_x \!+\! \cos k_y ) \!-\! 2i D' ( \cos k_x \!-\! \cos k_y ) . \nn
&
\label{eq:AkBk} \ea
Using abbreviations $X=\cos k_x, Y = \cos k_y$, we obtain the magnon energy spectrum
\begin{widetext}
\ba E_k = \sqrt{A_k^2 - |B_k |^2}= \sqrt{ [4J' \!-\! 2J (X \!+\! Y ) \!+\! Bn_0 ] [ 2J' (2 \!+\! X \!+\! Y) \!+\! Bn_0 ]  \!-\! (2D' )^2 (X\!-\!Y)^2 } . \label{eq:E-k} \ea
\end{widetext}
The spectrum has two local minima, at $k = 0$ and $k=Q=(\pi,\pi)$, with the minimum energy at $k=0$ given by

\ba E_0 &=& \sqrt{(4(J' -J) + Bn_0 ][8J' + Bn_0 ] } . \label{eq:2.31} \ea
It is governed by the larger of the DM energy $J'-J \sim D^2/J$ and the Zeeman energy $Bn_0$. Spin-rotation invariance of the Hamiltonian is completely lost due to the DM vector, and one sees a magnon gap of order $D^2/J$ even in the absence of Zeeman field.

The magnon spectrum derived from the Hamiltonian (\ref{eq:HDM}) is well-known~\cite{coffey90}, but little attention has been paid to the magnon eigenstates and the associated Berry curvature. The magnon eigenstate is given in the spinor form
\ba &&  |\psi_k \rangle = \bpm \cosh \theta_k /2 \\ -e^{-i \phi_k } \sinh \theta_k /2 \epm , ~~ e^{i \phi_k } = {B_k \over |B_k | } , \nn
& & \cosh {\theta_k \over 2} \!=\! \sqrt{ {A_k \over 2E_k } \!+\! {1\over 2} } , ~  \sinh {\theta_k \over 2} \!=\! \sqrt{ {A_k \over 2E_k } \!-\! {1\over 2} } .  \label{eq:1.19} \ea
Transformation to the quasiparticle operator $\gamma_k$ is implemented by the formula
\ba b_k = \cosh{\theta_k \over 2} \gamma_k - e^{i\phi_k} \sinh {\theta_k \over 2} \gamma^\dag_{-k} . \label{eq:bogoliubov} \ea
The Berry curvature of the magnon band can be calculated exactly~\cite{murakami11} as ($\partial_\mu = \partial/\partial k_\mu$, $\sigma_3$=Pauli matrix)
\ba {\cal B}_{k} &=&  i\langle \partial_x \psi_{k} | \sigma_3 | \partial_y \psi_{k} \rangle - i \langle \partial_y \psi_{k}  | \sigma_3 |  \partial_x \psi_{k}\rangle  \nn
&=& {1\over 2} \sinh \theta_{k} (\partial_x \phi_{k} \partial_y \theta_{k} - \partial_y \phi_{k} \partial_x \theta_{k} )   \nn
&=& 2 D' (J' \!+ \! J) (4J' \!+\! Bn_0 ) { \sin k_x \sin k_y \over E_k^3 } . \label{eq:2.35}\ea
The proportionality ${\cal B}_k \propto D'$ implies that the Berry curvature is possible only by the DM interaction. In the vicinity of $k=0$, one can write approximately
\ba {\cal B}_k \approx {16 D' J^2 \over E_0^3 } k_x k_y , \label{eq:Bk} \ea
which highlights the $d_{xy}$ character in the curvature function.

The thermal Hall conductivity $\kappa_{xy}$ is deduced from the Berry curvature through the formula developed by Murakami and collaborators~\cite{murakami11,murakami14}
\ba { \kappa_{xy} \over T} = - { k_B^2 \over \hbar } \int {d^2 k \over (2\pi)^2 } \left( c_2 ( E_k ) - {\pi^2 \over 3} \right) {\cal B}_k \label{eq:4.1} \ea
where $c_2 ( E_k )$ is some generalized Bose-Einstein distribution function of magnons. We find $\kappa_{xy} = 0$ by symmetry of the integral in (\ref{eq:4.1}). Specifically, ${\cal B}_k$ changes sign under either $k_x \rightarrow -k_x$ or $k_y \rightarrow -k_y$, but $E_k$ does not.

\section{Local defect scenario}

\subsection{Local spirals}
In a series of papers, Sushkov and collaborators have argued that holes introduced by doping Sr atom at the La site, for instance, get localized and distort the local spin configuration into a spiral with the wavevector $K =\sqrt{2} x (\pi, -\pi)$ for a given doping concentration $0 < x \lesssim 0.055$\cite{sushkov1,sushkov2,sushkov3}. For $0.055\lesssim x \lesssim 0.12$ the $K$ vector is directed along the crystallographic axis in accordance with the stripe scenario: $K = 2x (\pm \pi, 0)$.

Inspired by this proposal, we generalize the ground state triad (\ref{eq:triad}) to incorporate the spiral structure by writing
\ba \v n_i &=& n_1 (-1)^i \hat{a}_i - n_0 \hat{z} ,  \nn
\bm \theta_i & = & (-1)^i \hat{b}_i  \nn
\bm \phi_i & = & n_0 (-1)^i \hat{a}_i + n_1 \hat{z} ,  \label{eq:general-triad}\ea
where the local orthonormal vectors $\hat{a}_i$ and $\hat{b}_i$ are now position-dependent:
\ba \hat{a}_i & = & \hat{a} \cos \theta_i + \hat{b} \sin \theta_i  \nn
\hat{b}_i  & = & \hat{b} \cos \theta_i - \hat{a} \sin \theta_i . \label{eq:theta}\ea
Having $\theta_i = 0$ irrespective of the site $i$ corresponds to the magnetic order considered previously. Having $\theta_i = K\cdot r_i$ with $|K |\ll 1$ corresponds to the uniform spiral of slow modulation. Sushkov's scenario corresponds to having a finite rotation angle $\theta_i$ only in the vicinity of the impurity site. We will first consider the uniform spiral and the effect it has on the magnon Hall effect. Local spiral scenario will be considered subsequently.

There is an immediate consequence of having a finite spiral rotation angle $\theta_i$. The inner product $\bm \theta_i \cdot \bm \phi_j$ and $\bm \phi_i \cdot \bm \theta_j$, previously equal to zero in the general magnon Hamiltonian (\ref{eq:4.5}), is now finite:
\ba \bm \theta_i \cdot \bm \phi_j = n_0 \sin (\theta_i - \theta_j ) = -\bm \phi_i \cdot \bm \theta_j . \ea
Note that this term is nonzero only if the uniform moment $n_0$ is present simultaneously. As a consequence, the Hamiltonian matrix $H^0_k$ in (\ref{eq:2.27}) and (\ref{eq:AkBk}) is modified to $H_k^0 + H_k^1$, where
\ba H_k^1 = -2 J n_0 (\sin K_x \sin k_x + \sin K_y \sin k_y ) \sigma_3. \label{eq:Hk1} \ea
This new piece of Hamiltonian creates a simple shift in the magnon spectrum $E_k \rightarrow E_k +\delta E_k$,
\ba \delta E_k = - 2J n_0 (\sin K_x \sin k_x + \sin K_y \sin k_y ) . \ea
This is reminiscent of the Doppler shift; magnons whose momentum is parallel (anti-parallel) to $K = (K_x , K_y)$ experience a red-shift (blue-shift) in energy.

Meanwhile, the magnon wave function (\ref{eq:1.19}) and the Berry curvature (\ref{eq:2.35}) obtained earlier remain unchanged. In particular the various energy factors in the wave function and the Berry curvature are still those of the unperturbed Hamiltonian, maintaining the symmetries $E (k_x , -k_y ) = E (-k_x , k_y ) = E (k_x , k_y)$. The new magnon energy $E_k + \delta E_k$ enters solely through the distribution function $c_2 (E_k +\delta E_k)$ of the thermal Hall conductivity formula (\ref{eq:4.1}), which undergoes correction

\ba \delta \kappa_{xy} &\propto& \sum_k {\partial c_2 [E_k ] \over \partial E_k} {\cal B}_k  \delta E_k \nn
& \propto &  \sum_k  {\partial c_2 [E_k ]  \over \partial E_k } {\cal B}_k  (\sin K_x \sin k_x \!+\! \sin K_y \sin k_y ) = 0 . \nn \ea
The first two terms in the sum, $\partial c_2 / \partial E_k$ and $B_k$, are even under the change $k \rightarrow -k$, while $\delta E_k$ is odd. As a result, the sum must be zero. The uniform spiral state fails to produce Hall effect.

Akin to the original Sushkov proposal, we now look into the influence of localized spirals on the thermal Hall transport of magnons. First of all, we lay down some general strategy for attacking such problem. The continuum language is more appropriate for dealing with problems that break translation symmetry, and we begin with a continuum form of the Hamiltonian $H_1$ introduced in (\ref{eq:Hk1}):
\ba H_1 & = & i J n_0 \sum_{\langle ij\rangle} \sin (\theta_i - \theta_j ) (b^\dag_j b_i - b^\dag_i b_j ) \nn
& \rightarrow &  i J n_0 \int_r \bm \nabla \theta \cdot (b^\dag \bm \nabla b - b \bm \nabla b^\dag ) . \ea
The integral symbol $\int_r = \int dx dy$ is abbreviated. Spatial gradient of the phase $\bm \nabla \theta$ is localized around the impurity site. The solution worked out by Sushkov {\it et al.} gives
\ba \theta_\alpha &=& f(|r - r_\alpha |) \hat{b} \cdot (r - r_\alpha ), \label{eq:4.5} \ea
around each impurity centered at $r_\alpha$. For a collection of impurities the phase twist is the sum $\theta = \sum_\alpha \theta_\alpha$. The envelope function $f(r)$ approaches a constant $f_0$ at the center of impurity and produces a spiral-like configuration locally.

At the level of Born scattering, the perturbation $H_1$ fails to produce any Hall-like transport of magnons. To see this one writes $H_1$ in Fourier space,
\ba
H_1 =  - i J S n_0 \sum_{k,p} p \cdot (p+ 2 k) \theta_p b^\dag_{k+p} b_k , \label{eq:H1} \ea
where $\theta_p$ is the Fourier transform of the real-space $\theta$. The Born scattering amplitude $\langle k+p |H_1 |k\rangle$ is proportional to the factor $p \cdot ( p+2k )  = (k+p)^2 - k^2$. Under the continuum approximation, however, the quasiparticle energy $E_k$ is a quadratic function of $k$ (see Eq. (\ref{eq:E-k}) for the full energy dispersion). Elastic scattering process satisfies $E_{k+p}= E_k$, hence $(k+p)^2 - k^2 = 0$. The Born scattering amplitude vanishes. Higher-order contributions from $H_1$ involves higher powers of the uniform moment $n_0$ and are expected to give negligible contribution.

Upon expanding to one higher order in the phase gradient, we do find an additional correction in the form
\ba
H_2 &\approx & {1\over 4} JS \sum_r (\bm \nabla \theta )^2 (b^\dag - b)^2 . \label{eq:H2} \ea
Born scattering calculation based on this Hamiltonian also gives negative results for the magnon Hall effect. Details are not illuminating and omitted from readership.

\subsection{Local skyrmions}

Speculations of skyrmion formulation in the cuprates have been around for a long time~\cite{shraiman,gooding,haas} and revived recently with the report of their sightings in a member of the cuprate family La$_2$Cu$_{0.97}$Li$_{0.03}$O$_4$~\cite{popovic}. It has been well-established in the recent skyrmion literature that a magnon sees a localized skyrmion as two units of flux quanta~\cite{zang13,nagaosa-review,zhao-book,jiang-review,han-book}, and will experience Aharonov-Bohm scattering. We examine whether such scenario can apply to the antiferromagnetic skyrmions, assuming they do form localized defects in the underdoped or undoped cuprates.

In a nutshell, an antiferromagnetic skyrmion {\it per se} does not give rise to magnon Hall effect, although the ferromagnetic skyrmion does. The difference can be outlined most simply in the continuum field theory of magnons for each case. For ferromagnetic model we switch $J \rightarrow -J$ in the magnon Hamiltonian and treat $\v n_i, \bm \theta_i, \bm \phi_i$ as smooth functions of the coordinates, as there is no staggered component in any of them. Continuum limit of the magnon Hamiltonian with $\v D_{ij} =0$ and $\v B =0$ is easily obtained as
\ba H_{\rm FM} \sim - {J\over 2} \sum_\mu b^\dag  [\partial_\mu - i a_\mu ]^2 b  + \cdots \ea
where the curl of the vector potential $\partial_x a_y - \partial_y a_x = (2\pi)^{-1} \v n \cdot (\partial_x \v n \times \partial_y \v n)$ represents the local skyrmion density. Integral of the curl $(\bm \nabla \times \v a)_z$ is -2 for a skyrmion charge -1. This is the basis of the claim that the local skyrmion magnetic structure acts as two units of flux quanta for the magnons. The magnon Hall effect due to skyrmion has been observed experimentally in ferromagnetic thin films~\cite{skyrmion-ratchet}.

A very different effective theory of magnons is found for antiferromagnetic ground states. The smooth texture is realized for the {\it staggered magnetization}, so the ground state triad is parameterized as
\ba \v n_i \rightarrow (-1)^i \v n_i^s , ~~ \bm \theta_i \rightarrow (-1)^i \bm \theta_i^s , ~~ \bm \phi_i \rightarrow \bm \phi_i . \ea
Both $\v n_i$ and $\bm \theta_i$ are staggered but not $\bm \phi_i$, which is defined as the cross product $\bm \phi_i = \v n_i \times \bm \theta_i$. Now treating $\v n^s_i$ and $\bm \theta^s_i$ and $\bm \phi_i$ as smooth, we obtain the continuum magnon Hamiltonian
\ba H_{\rm AFM}=  {J \over 2} \sum_\mu \bigl[  (\partial_\mu b^\dag)^2 + (\partial_\mu b )^2 \bigr]  + \cdots. \ea
Various other terms proportional to $b^2$, $(b^\dag )^2$ and $bb^\dag$ are not shown. Crucially, there is no analogue of the covariant derivative $\partial_\mu - i a_\mu$ in this theory and no source of emergent magnetic field. Magnon Hall effect originating from skyrmion spin texture must be absent in the antiferromagnetic ground state.

\begin{figure}[htbp]
\includegraphics[width=0.3\textwidth]{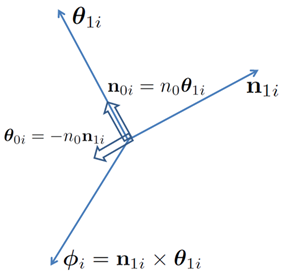}
\caption{Coordinate system used in developing the magnon dynamics of ferrimagnetic ground state}
\label{fig:1}
\end{figure}

As we saw earlier, however, undoped cuprate is weakly ferrimagnetic, due to the DM interaction and the consequent canting of spins. Since ferrimagnet has characters of both ferromagnet and antiferromagnet, we find it worth exploring possible low-energy magnon dynamics for a ferrimagnetic spin-textured ground state. To this end one needs a more elaborate setup for treating magnon dynamics by allowing the triad of orthonormal vectors $(\v n_i, \bm \theta_i, \bm \phi_i )$ to carry both staggered and uniform components locally:
\ba \v n_i &=&  (-1)^i \v n_{1i}  + n_0 \bm \theta_{1i}  , \nn
 \bm \theta_i & = & (-1)^i \bm \theta_{1i} - n_0 \v n_{1i}  ,  \nn
 \bm \phi_i & = & \v n_{1i} \times \bm \theta_{1i} .  \label{eq:ferri2} \ea
Words of explanation are in order for this choice of parametrization. The uniform moment $\v n_{0i}$ is by assumption orthogonal to the staggered moment $\v n_{1i}$. The staggered component of $\bm \theta_i$, denoted $\bm \theta_{1i}$, is also orthogonal to $\v n_{1i}$. Since both $\v n_{0i}$ and $\bm \theta_i$ are required to be orthogonal to $\v n_{1i}$, and there is a U(1) degree of freedom in choosing the orthonormal vector $\bm \theta_{1i}$, we invoke this freedom to choose $\bm \theta_{1i}$ to be parallel to $\v n_{0i}$, or write $\v n_{0i} = n_0 \bm \theta_{1i}$. This explains the parameterization of $\v n_i$ in the first line of (\ref{eq:ferri2}). The second line for $\bm \theta_i$ follows naturally from requiring $\v n_i \cdot \bm \theta_i = 0$. Orthogonality of all three vectors in (\ref{eq:ferri2}) is ensured up to first order in the small moment $n_0$. The parameterization we propose is summed up pictorially in Fig. \ref{fig:1}.

Substituting (\ref{eq:ferri2}) into the general magnon Hamiltonian (\ref{eq:4.5}) yields terms, linear in $n_0$,
\ba \bm \theta_i \cdot \bm \phi_j &\approx& -n_0 \v n_{1i} \cdot (\v n_{1j} \times \bm \theta_{1j} )  \rightarrow - \v n_0 \cdot (\v n_1 \times \partial_\mu \v n_1 ) , \nn
\bm \phi_i \cdot \bm \theta_j &\approx& - n_0 \v n_{1i} \times \bm \theta_{1i}  \cdot \v n_{1j} \rightarrow \v n_0 \cdot (\v n_1 \times \partial_\mu \v n_1 )  .
\ea
In arriving at the expressions at the far right we assumed continuum approximation and introduced $\partial_\mu$ for the spatial derivative in the direction $j = i +\hat{\mu}$. A new
contribution to the magnon dynamics arises from
\ba H_1 &=&  i J  \sum_{\mu = x, y}  \int dx dy ~ (\v n_0 \cdot \v n_1 \!\times \!\partial_\mu \v n_1 ) (b^\dag \partial_\mu b \!-\! b \partial_\mu b^\dag )  \nn
& = & J \sum_{\mu =x,y} \int dx dy ~ \v a \cdot \v j , \ea
where the vector potential $\v a$ and the magnon current density $\v j$ are defined by
$a_\mu = - \v n_0 \cdot \v n_1 \!\times \!\partial_\mu \v n_1$, and
$j_\mu  = -i (b^\dag \partial_\mu b - b \partial_\mu b^\dag )$, respectively.

For non-textured ground state, the uniform and staggered moments are related by  $\v n_0 = n_0 \hat{b} \times \v n_1$ through the DM interaction. If we assume that this relation continues to hold even for the textured spin configuration such as that of a skyrmion, it turns out one can write the vector potential in a much simpler form: $\v a = - n_0 \bm \nabla (\hat{b} \cdot \v n_1 )$. In this case, the Hamiltonian $H_1$ reduces exactly to the form $H_1 \sim J n_0 \bm \nabla \theta \cdot \v j$ we discussed in the earlier subsection. The Born scattering amplitude there was zero, and so is it here. To conclude, even the ferrimagnetic skyrmion scenario fails to produce skew scattering at the level of Born scattering. Again, more elaborate theories are likely to involve higher powers of $n_0$ and very small.

All of the local defect scenarios considered in this section fail to show skew scattering, at least at the lowest order in the uniform moment $n_0$. There is also a general issue how to reconcile the impurity-induced defects with the undoped cuprate, where the impurities are nominally absent. Finally, the magnon gap grows with the magnetic field and suppresses the response function in any magnon-based scenarios. The experiment on $\kappa_{xy}$ does not show such activation behavior~\cite{taillefer}.

\section{Fermionic spinon theory of thermal Hall effect}

With the general inability of the magnon theory to account for the observed thermal Hall effect in the undoped to lightly doped cuprates, we turn to look for an alternative theory.  A very natural candidate is to assume the existence of spinon excitations. There are many different classes of spinon models~\cite{wen02}. Within the context of the cuprates, a common starting point is the so-called $\pi$-flux phase, where there is $\pi$ flux per plaquette resulting in fermion spinons with a Dirac dispersion at $(\pi /2,\pi /2)$ and symmetry-related points. It is then assumed the due to strong gauge field fluctuations, the spinons are bound in a confined phase and antiferromagnetism appears, so that the only low energy fluctuations are $S=1$ spin waves (for a review see Ref.~\onlinecite{kim_lee}) The spinon idea had been adopted also to compute spin dynamics in the undoped cuprates~\cite{hsu90,ho01}, even though long-range magnetic ordering in such compound was well established. The spinon-based theories were rationalized by the fact that some aspects of high-energy spin excitations are not captured by the spin-wave picture alone, and that a vestige of spinon excitations must remain in the physical spectrum to account for the spin dynamics fully. However, the expectation has been that the spinon gap is relative large (a fraction of $J$) and that the confined spinons will not influence low temperature properties. Hence the spinon-based theories have not been applied to low-energy transport properties such as the thermal Hall conduction. The recent experiment, taken at its face value, calls for a re-evaluation of this traditional view.

Historically the spinons are discussed in the context of the spin liquid state, where there is no antiferromagnetic order. However, it has been pointed out that this restriction is unnecessary, and there is a possibility of spinon excitations co-existing with antiferromagnetic (AF) order and spin waves. Such a state possesses topological order and has been called AF*. This scenario was first proposed by Balents, Fisher and Nayak~\cite{BFN99} and have been further discussed by Senthil and Fisher~\cite{SF01}.  They started with a $d$-wave superconductor which they disordered by  proliferating  $hc/e$ vortices while the $hc/2e$ vortices remain gapped. The resulting state is called a nodal spin liquid with Dirac spinons that grew out of the $d$-wave Bogoliubov quasi-particles. This state can co-exist with antiferromagnetism at wavevector $(\pi,\pi)$. If the nodes are connected by the AF wavevector, they will be gapped. On the other hand, if their separation is not $(\pi,\pi)$ they may remain as gapless Dirac fermions. Guided by this line of thinking, it seems to us that the next step is to start with a Dirac spectrum given by the $\pi$ flux model and simply assume that it co-exists with the antiferromagnetism. 

We proceed to first present a simple spinon-based model of magnetic dynamics, and use it to compute thermal Hall conductivity and spin chirality. Our goal is to find the simplest model that can give result that are qualitatively similar to the observed thermal Hall effect. Even this is a highly nontrivial task because the experiment imposes serious constraints. First, as we shall see, the observed $\kappa_{xy}^{\rm 2D}/T$ is very large when expressed in the natural unit of $k_B^2 /\hbar$ per layer. Second, the effect is seen down to quite low temperature of about 5K, which says that the spinon gap cannot be too large. Third, the effect is linear in B. This rules out chiral spin liquid states which spontaneously break time reversal symmetry and which will lead to hysteretic behavior that is not seen experimentally. While the Dirac nodes have Berry curvatures near the nodes, they are canceled between two sets of nodes and by itself will not give rise to thermal Hall effect. Thus we need to introduce some chirality which is induced linearly with the applied magnetic field. There are two ways an external field couples to the spinons. First is via Zeeman effect and the second is a coupling to the spin chirality, As we shall discuss later, this latter coupling is proportional to the flux generated by the external field per plaquette and is extremely small. Therefore we consider the Zeeman coupling only and we come up with the model discussed below. We do not think this model is realistic for the Cuprates. We assume a spin dependent hopping which is possible only in the presence of spin-orbit coupling, which is not believed to be strong for the Cuprates. At this stage of the development, we believe there is value in this exercise, if only to emphasize the challenge we face in coming up with even a phenomenological model that satisfy the experimental constraint outlined above.

We outline general requirements in a candidate spinon model. Firstly, it will consist of spin-up and spin-down fermion bands with identical dispersions and opposite Berry curvatures. As such, the Hall effect of one species of fermions will be cancelled out by that of the other. The applied magnetic field will then split the energy degeneracy and lead to the non-cancellation of Berry curvatures, resulting in non-zero thermal Hall conductivity. In such picture, the predicted Hall signal will be naturally proportional to the field strength $B$: $\kappa_{xy} \propto B$ - a prominent feature in the observed thermal Hall effect in underdoped cuprates~\cite{taillefer}.

\begin{figure}[htbp]
\includegraphics[width=0.45\textwidth]{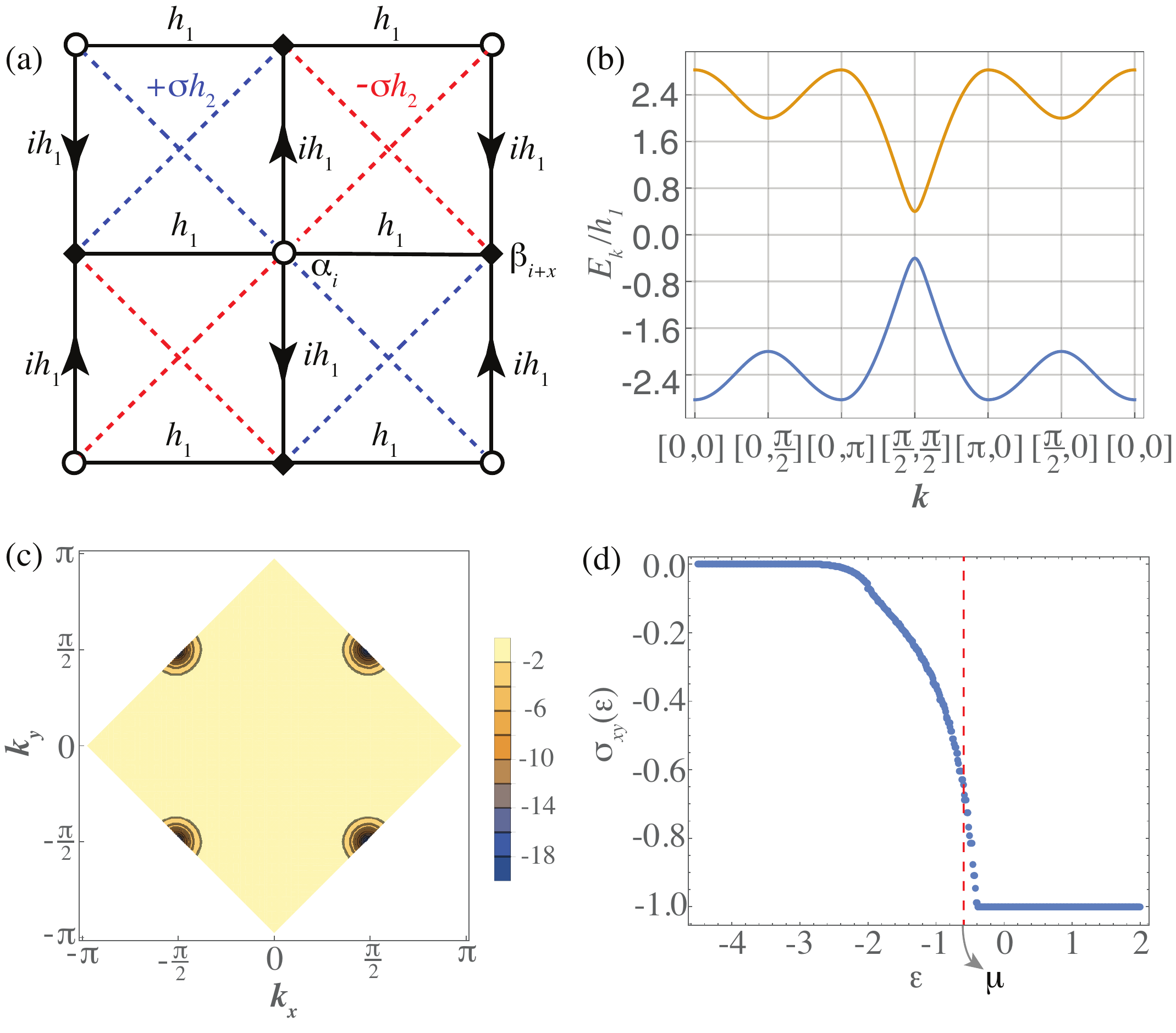}
\caption{(a) Hopping parameters adopted in our fermion model. Arrows indicate the imaginary hopping direction. Sign of the diagonal hopping depends on spin $\sigma = \pm 1$. (b) Upper and lower band dispersions obtained from the spinon model. (c) Berry curvature of the lower band over the Brillouin zone. (d) Hall conductivity $\sigma_{xy} (\epsilon)$. States with $\epsilon < \mu$ are occupied at zero temperature. Plots are drawn with $h_1 = 1, h_2 =0.1$.}
\label{fig:2}
\end{figure}

The model we present can be summed up as a $2\times2$ fermion Hamiltonian
\ba H = {1\over 2} \sum_{k\sigma} \psi^\dag_{k\sigma} H_{k\sigma} \psi_{k\sigma} . \label{eq:sachdev-H} \ea
For each spin $\sigma=\uparrow,\downarrow$ we have the spinor $\psi_{k\sigma} = \bpm \alpha_{k\sigma} \\ \beta_{k\sigma}\epm$, and the Hamiltonian matrix
\ba H_{k\sigma} =\bpm 4 \sigma h_2 s_x s_y - \sigma B & 2h_1 ( c_x + i c_y ) \\  2 h_1 ( c_x - i c_y ) & - 4 \sigma h_2 s_x s_y - \sigma B  \epm . \nonumber \\ \ea
We have used the abbreviations $c_{x (y)} = \cos k_{x (y)} , s_{x (y)} = \sin k_{x (y)}$.The hopping amplitudes in the nearest-neighbor and the diagonal directions are as displayed in Fig.~\ref{fig:2}. Without the diagonal hopping this is the $\pi$-flux Hamiltonian whose energy spectrum has Dirac nodes~\cite{hsu90,ho01}. The diagonal hopping term $h_2$ opens up a gap at the Dirac points and creates bands with Chern numbers. The spin-dependent diagonal hopping amplitude is designed to generate opposite signs of Berry curvature between the two spin orientations. The Zeeman energy $-\sigma B$ is included in the Hamiltonian.

Diagonalizing the Hamiltonian, we find the energy and the Berry curvature of the bands:
\ba E_{n k\sigma} & = & 2 n \left( h_1^2 ( c_x^2 + c_y^2 ) + 4 h_2^2 s_x^2 s_y^2 \right)^{1/2} - \sigma B , \nn
{\cal B}_{n k \sigma} & = & 2 n \sigma  {h_1^2 h_2 (  1- c_x^2 c_y^2 )  \over \left( h_1^2 ( c_x^2 + c_y^2 ) + 4 h_2^2 s_x^2 s_y^2 \right)^{3/2} } . \label{eq:EandB}  \ea
The band index $n = \pm 1$ refers to the upper and the lower band, respectively. The Berry curvature ${\cal B}_{nk\sigma}$ has opposite signs between the two bands, and between the two spins. For visualization of the band dispersion and the Berry curvature, see Fig. \ref{fig:2}. The upper and lower bands are separated by a gap of magnitude $8 |h_2 |$ at $(k_x , k_y ) = (\pm \pi/2, \pm \pi/2)$.


The zero-temperature Hall conductivity at the putative chemical potential $\epsilon$ for each spin species is derived from the Berry curvature through the TKNN formula~\cite{thouless82}
\ba & \sigma_{xy\sigma} (\epsilon) = \sum_{n k} {\cal B}_{n k\sigma} \theta (\epsilon - E_{nk\sigma} )  ,   \ea
which involves the sum over all states whose energies lie below $\epsilon$. In the quantized case we obtain $\sigma_{xy} = C/2\pi$, where $C$ is the Chern number. The lower band in our fermion model has the spin-dependent Chern number $C_\sigma = - \sigma$ for $\sigma = \pm 1$ ($\uparrow, \downarrow$). For calculation of thermal conductivity in the fermionic model we use the formula derived in Ref. \cite{niu11},
\ba { \kappa^{\rm 2D}_{xy} \over T} ={1\over 4 T^3} \int d\epsilon { (\epsilon - \mu )^2 \over \cosh^2 [\beta (\epsilon-\mu)/2 ]  } \sigma^{tot}_{xy} (\epsilon) . \label{eq:kappa-formula} \ea
This has the form of a well-known Mott formula relating the thermal conductivity to the electric conductivity. To restore physical units to the dimensionless form of $\kappa^{\rm 2D}_{xy}/T$ given above, one has to multiply by $k_B^2/\hbar$, the ratio of Boltzmann's constant and the Planck's constant. It is useful to note that $k_B^2/\hbar = 1.81\times 10^{-12}$ W/K$^{2}$. As an example, consider a bulk La$_2$CuO$_4$ sample whose $c$-axis constant is $d=13.2\AA$. Since there are two CuO$_2$ layer per unit cell, the effective inter-layer distance is half that, $d_{\rm eff} = 6.6 \AA$. If each CuO$_2$ layer carried a two-dimensional $\kappa^{\rm 2D}_{xy}/T$ worth the universal value $k_B^2/\hbar$, the three-dimensional thermal Hall conductivity of the bulk La$_2$CuO$_4$ would be given by $\kappa_{xy}^{\rm 3D}/T = \kappa_{xy}^{\rm 2D}/(T \cdot d_{\rm eff} )= 2.76$ mW/K$^2$m. The recently observed thermal Hall conductivity in cuprates reaches maximal $\kappa_{xy}^{\rm 3D}$ values in the vicinity of 30-40 mW/Km at $T\approx 10$K, consistent with a per layer value of  $\kappa_{xy}^{\rm 2D}/T$ roughly equal to $k_B^2/\hbar$ at that temperature. The thermal Hall conductivity formula (\ref{eq:kappa-formula}) predicts values of $\kappa_{xy}^{\rm 2D}/T$ in the range of $k_B^2 /\hbar$ for $\sigma_{xy} \sim 1$.

\begin{figure}[htbp]
\includegraphics[width=0.4\textwidth]{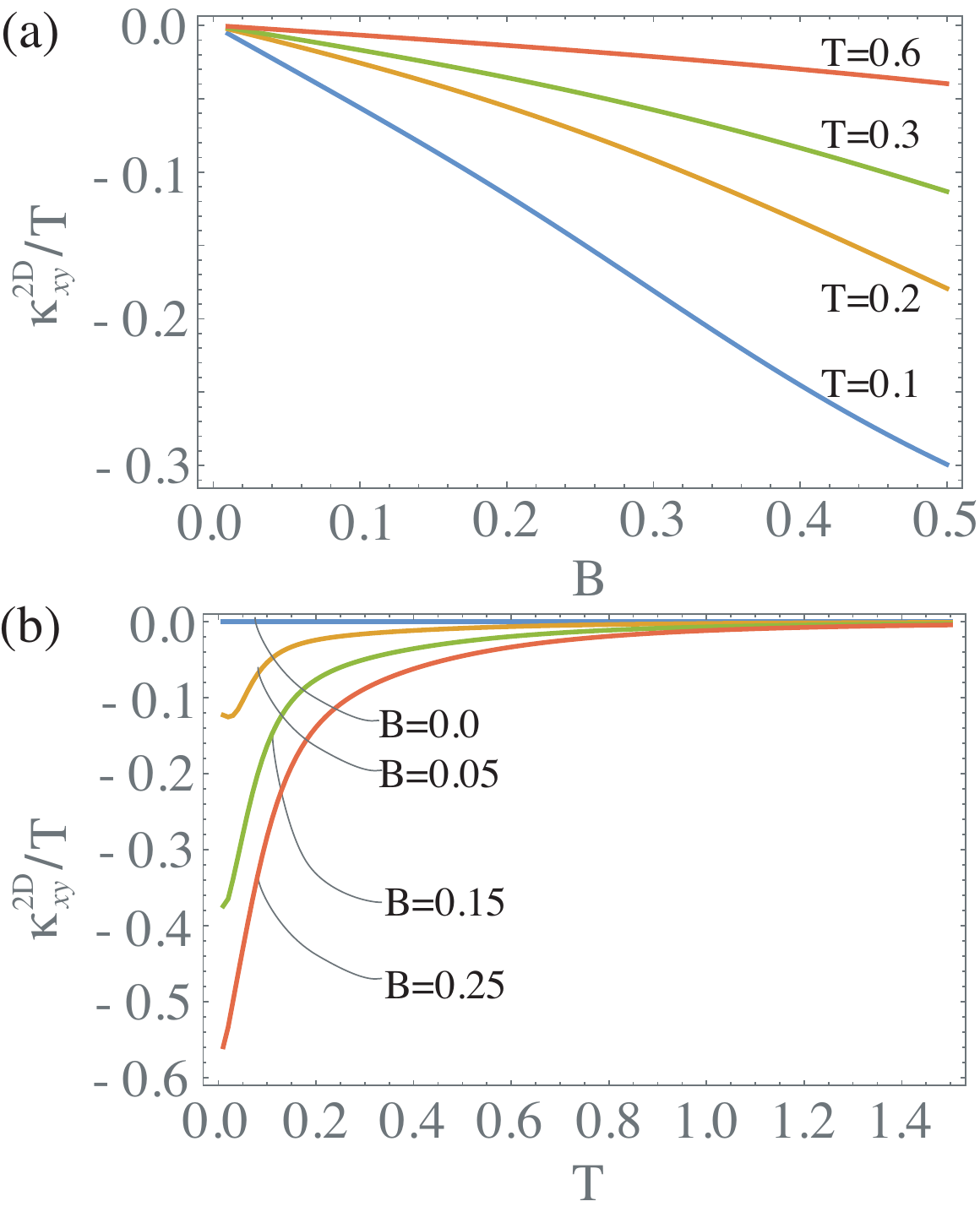}
\caption{Two-dimensional thermal Hall conductivity $\kappa^{\rm 2D}_{xy}/T$ (in physical units of $k_B^2/\hbar$) vs. (a) magnetic field $B$ at several temperatures $T$ and (b) vs. temperature $T$ at several magnetic fields $B$. Parameters chosen are $h_1 = 1, h_2 = 0.1$, and the chemical potential $\mu = -0.6$, corresponding to the filling factor $n=0.98$ at $T=0, B=0$. Temperature and magnetic field scales are measured in units of $h_1$. }
\label{fig:3}
\end{figure}

The Hall conductivity $\sigma_{xy}^{tot}$ itself is given as the sum of contributions from the two spin species: $\sigma^{tot}_{xy} (\epsilon) = \sigma_{xy, \uparrow} (\epsilon) + \sigma_{xy, \downarrow} (\epsilon)$. In the absence of Zeeman field we have the opposite signs of the Berry curvature and the degenerate energy bands, i.e. ${\cal B}_{nk\uparrow} = - {\cal B}_{nk\downarrow}$ and $E_{nk\uparrow} = E_{nk\downarrow}$, hence a vanishing Hall conductivity: $\sigma_{xy, \uparrow} (\epsilon) + \sigma_{xy, \downarrow} (\epsilon) = 0$. The energy degeneracy of $\uparrow,\downarrow$-spinons are split by the Zeeman field, whereas the Berry curvature itself remains unaffected by it. The Hall conductivity formula in the presence of $B$ becomes
\ba \sigma_{xy}^{tot} (\epsilon ) = \sigma_{xy} (\epsilon +B ) - \sigma_{xy} (\epsilon- B )  . \label{eq:Pauli-para}\ea
Here $\sigma_{xy} (\epsilon) = \sigma_{xy, \uparrow} (\epsilon)$ is the Hall conductivity of $\uparrow$-spinons. There is more occupation of $\uparrow$-spinons than $\downarrow$-spinons, because the chemical potential for the former (latter) particle has been raised (lowered) by $B$.
In the model Hamiltonian we chose, the $\uparrow$-spinon band carries the Chern number $-1$ and results in negative values of $\kappa_{xy}^{\rm 2D}$.

Numerical calculation of the thermal Hall conductivity as a function of temperature and magnetic fields are shown in Fig. \ref{fig:3}. The chemical potential was chosen in such a way that the average occupation number was $\langle f^\dag_{i\sigma} f_{i\sigma} \rangle = n$ at zero temperature and magnetic field. The linear-$B$ dependence of $\kappa^{\rm 2D}_{xy}/T$ in the numerical plot is easy to understand, since  $\sigma_{xy} (\epsilon + B) - \sigma_{xy} (\epsilon - B)  \propto B$ at small values of $B$. Thermal smearing reduces the Hall signal at higher temperatures. The magnitude of $\kappa_{xy}^{\rm 2D}/T$ values calculated within our model can reach values close to one ($k_B^2/\hbar$ in physical units) with suitable choices of $h_2$ and $\mu$.

The spinon density $n$ was chosen to be 0.98 in the calculation of thermal Hall conductivity, Fig. \ref{fig:3}. In the slave fermion model the spinon density equals the electron density on average, and at the Mott insulator limit $n$ should be unity. In our model for $n=1$ the thermal Hall effect is zero at zero temperature because the chemical potential will lie in the gap. However, it will be finite for sufficiently large B and/or temperature. The value 0.98 may be considered slightly doped. Results for other values of $n$ will be shown later.

The fermion model we study supports the spin chirality as well. In the mean-field theory, average of the spin-chirality operator $\v S_i \cdot (\v S_j \times \v S_k)$ of the $\langle ijk \rangle$ triangle becomes, through the substitution $\v S_i = (1/2) f^\dag_i \bm \sigma f_i$ with $f_i = (f_{i\uparrow} ~ f_{i\downarrow})$,
\ba  \langle \v S_i \cdot (\v S_j \times \v S_k ) \rangle = -{i \over 2} \Bigl(  \chi_{ij} \chi_{jk} \chi_{ki} -  \chi_{ik} \chi_{kj} \chi_{ji} \Bigr)  , ~~~ \ea
where $\chi_{ij} = \sum_\sigma \langle f^\dag_{i\sigma} f_{j\sigma} \rangle$. Calculations of $\chi_{ij}$ in the mean-field theory is straightforward. The essential point, as it turns out, is that the triple product of hopping parameters $\chi_{ijk} \equiv \chi_{ij}\chi_{jk}\chi_{ki}$ contains an imaginary term only at finite magnetic field and diagonal hopping, thus $\langle \v S_i \cdot (\v S_j \times \v S_k ) \rangle = {\rm Im} [ \chi_{ijk} ]\propto h_2 \cdot B$.
%
%
%
%

\begin{figure}[htbp]
\includegraphics[width=0.4\textwidth]{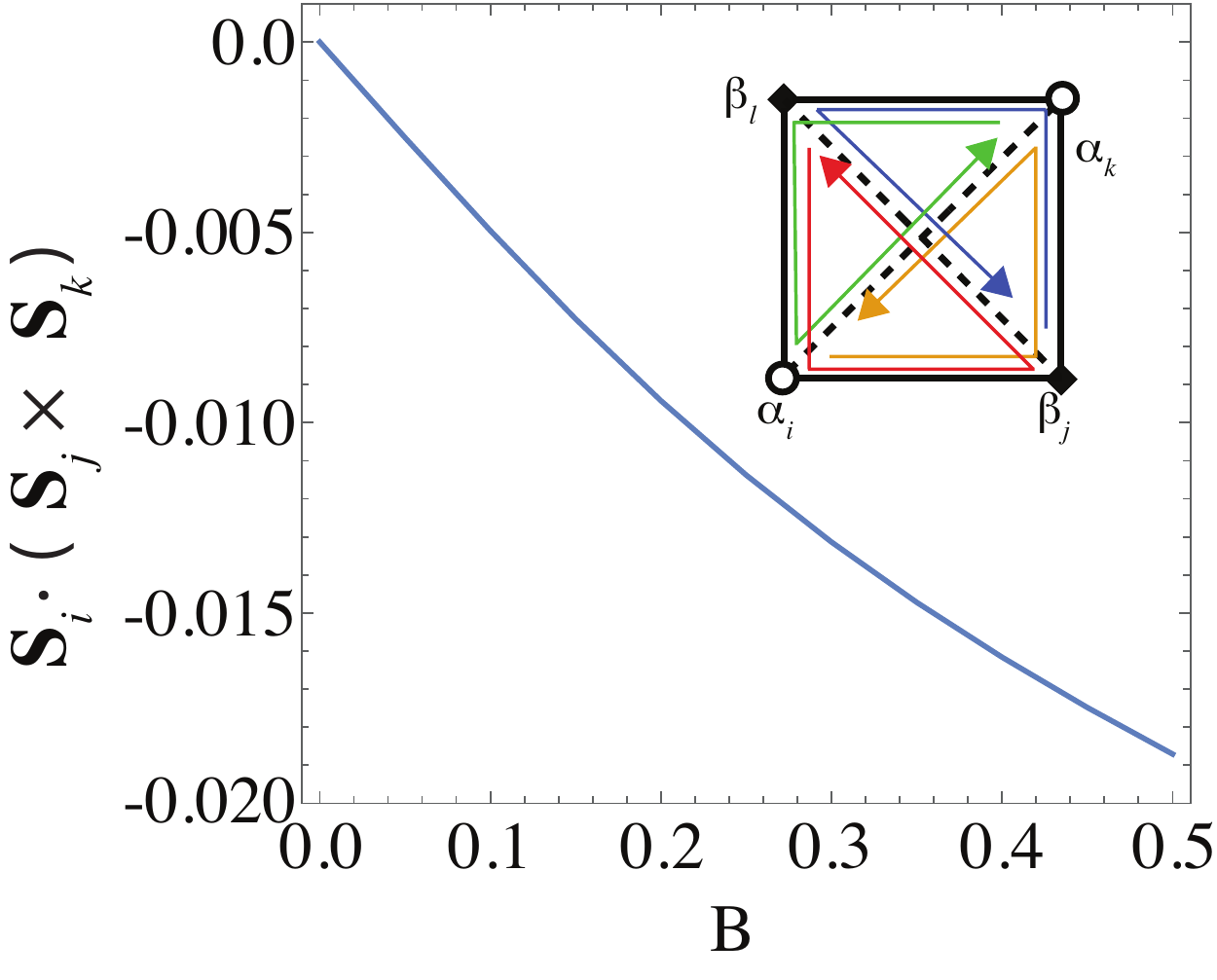}
\caption{Magnetic field dependence of spin chirality $\langle \v S_i \cdot (\v S_j \times \v S_k ) \rangle$ for the triangles of the elementary square. It grows linearly with $B$ at small fields. Parameters used are $h_1 =1.0, h_2 = 0.1$, and $\mu = -0.6$ $(n=0.98)$ as in Fig. \ref{fig:3}. (inset) four corners of the elementary square are labeled by $i,j,k,l$. Spin chirality is calculated for each of the four triangles by going in the counter-clockwise fashion. All four triangles carry the same value of spin chirality.  }
\label{fig:4}
\end{figure}

Explicit calculation shows all elementary triangles having the same spin chirality. In other words, finite magnetic field induces uniform spin chirality state within our model. Numerical evaluation of spin chiralities through the four triangles of the elementary square are shown in Fig. \ref{fig:4}, displaying the expected linear growth with $B$ at small fields.
Our observation suggests that an interaction of the form $\sim B \v S_i \cdot (\v S_j \times \v S_k )$ might be present and play a hitherto neglected role in the transport of undoped cuprates. Such interaction Hamiltonian is well-known to derive from the large-$U$ expansion of the Hubbard interaction, when an external magnetic field is present~\cite{sen95}. Application of such spin chirality Hamiltonian to the understanding of the behavior of spin liquid phase under external magnetic field was taken up in Ref.~\cite{motrunich06}, where the focus had been  the orbital effects of the magnetic field such as the Landau level formation of spinons, without explicit consideration of the Zeeman splitting of the spinons as we do. The spinon hopping parameters in Ref.~\cite{motrunich06} pick up an imaginary part as a result of the Aharonov-Bohm effect, while our hopping parameters are deemed fixed and unchanged under the magnetic field. We also note that a spin chirality induced by a magnetic field was considered earlier by Katsura {\it et al.}~\cite{KNL} to generate a thermal Hall effect. However that effect is extrinsic, {\it i.e.} it depends on the scattering of the spinons by disorder, whereas the effect we consider in this paper is intrinsic.

\begin{figure}[htbp]
\includegraphics[width=0.45\textwidth]{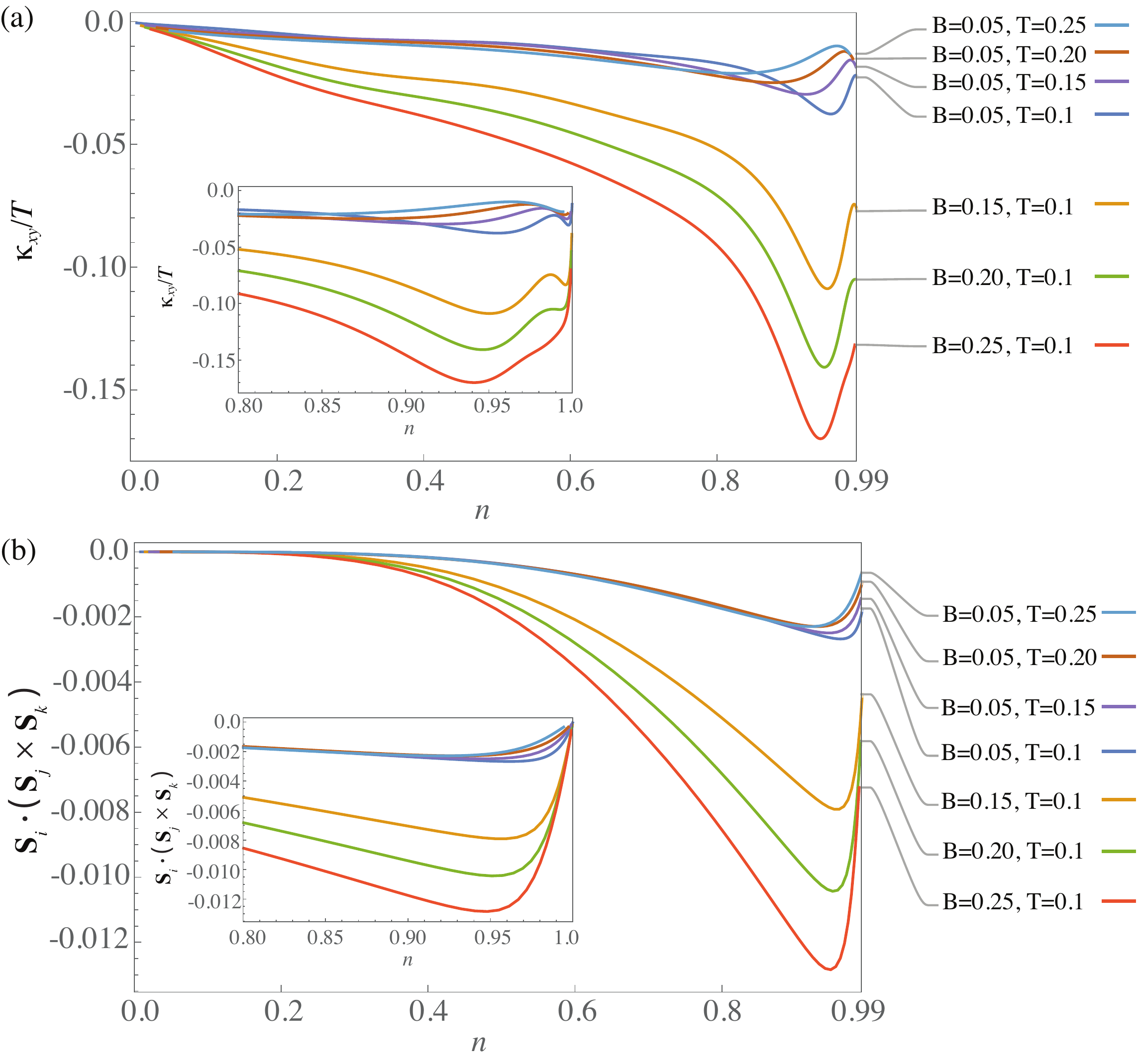}
\caption{Doping ($n$) dependence of (a) $\kappa_{xy}/T$ and (b) spin chirality at several values of $T$ and $B$. }
\label{fig:5}
\end{figure}

Figure \ref{fig:5} shows the doping dependence of $\kappa_{xy}$ and spin chirality at some fixed temperature and field. As one can see, the $\kappa_{xy}/T$ reaches a maximum in the vicinity of $n\approx 0.95$ in our model. The spin chirality nearly vanishes at $n=1$, since the two orientations of spinons actually carry opposing sense of circulation, i.e. $\chi_{ij, \uparrow} \chi_{jk,\uparrow} \chi_{ki, \uparrow} \approx - \chi_{ij, \downarrow} \chi_{jk, \downarrow}\chi_{ki,\downarrow}$, and it is the residual part of their sum which contribute to the spin chirality. At $n=1$ the cancellation is almost complete, hence the spin chirality becomes very small.
Additionally, one can check that spin-spin correlation $\langle \v S_i \cdot \v S_j\rangle$ preserves the lattice symmetries as well, and the loss of translational and rotational symmetry in the hopping patterns of our ansatz is only an artifact of the spinon theory. The aspect of projective symmetry restoration was discussed in Ref.~\cite{sachdev18} also.


The spinon model we propose is not without its drawbacks. On the theoretical side, the conventional view is that starting from a spinon model, the N\'{e}el state can emerge as a confinement transition, where the spinons become gapped and confined~\cite{kim_lee}. Thus we normally do not expect the co-existence of antiferromagnetic order and  nearly free spinons. On the other hand, such co-existence is allowed but should be considered highly exotic~\cite{BFN99,SF01}. Furthermore, the spinon gap must be small in the insulator in order to give a thermal Hall effect at relatively low temperature and magnetic field. The particular spinon dynamics that we assume, with spin-dependent hopping, does not have a well-defined microscopic justification at the moment, except that it might in some way be tied to spin-orbit interaction. The model on the whole is an attempt to fit the observation.  On the experimental side, the renormalized spin-wave theory does a good job in accounting for the magnetic excitations in the square-lattice antiferromagnet, as revealed for instance in recent experiments~\cite{ronnow12,ronnow16}. On the other hand, some high-energy features in the magnetic excitation are not fully explained within the spin-wave theory alone~\cite{ronnow12,ronnow16}, which in turn prompted speculations about residual spinon excitations in the Heisenberg model~\cite{sandvik17}. Overall it is fair to say that at this point, spinons as low-energy excitations in square-lattice quantum antiferromagnet has quite weak experimental support. On the other hand, the two quasiparticles - magnons and spinons - give contrasting predictions in regard to their behavior under the magnetic field. In the spin-wave scenario, a magnon gap inevitably opens and suppresses magnon contribution to transport. For the spinon-based scenario, as demonstrated here, linear growth of the response function $\kappa_{xy}/T$ with the field is natural. The diagonal spinon hopping term $\sim h_2$ necessary for the opening of the gap, the existence of Berry curvature, and ultimately the thermal Hall transport, all seem closely related to the spin chirality correlation, given that the latter quantity scales with $h_2$ in our model. In turn, including the three-spin exchange interaction on top of the Heisenberg interaction might be a necessary ingredient for the complete understanding of magnetic dynamics in undoped cuprates.

If the spinon excitations indeed play a role in the thermal transport in the antiferromagnetic phase of the cuprates, they must have manifestations on other probes such as inelastic neutron scattering and heat capacity measurement. Calculations of such physical quantities within the same spinon scenario, coupled with critical re-examination of past experiments in light of such theory, might shed further light on the true nature of low-energy excitations in the undoped cuprates. Thermal Hall measurement on other square-lattice antiferromagnets will be a nice cross-check on the observed effect in the cuprates as well. 
\\

{\it Note added:} Spinon theory of thermal Hall effect in magnets with Dzyaloshinskii-Moriya interaction was also advanced in a recent preprint~\cite{chen19} and applied to the Kagome lattice. We also mention a preprint by Chatterjee {\it et al.}~\cite{chat19} which also used the $\pi$ flux spinon as a starting point. A key ingredient is the term $J_\chi \sum_\triangle \v S_i \cdot ( \v S_j \times \v S_k )$ in their Eq. (2), where $J_\chi$ is proportional to the magnetic flux through a triangular plaquette. This term generates a net chirality which produces a thermal Hall effect. We had considered this term in the last section but did not discuss it further because of the very small magnitude. One can make an estimate of $J_\chi$ using the $t/U$ expansion by Motrunich~\cite{motrunich06}, to find $J_\chi = -48 \pi (t_2 t^2 / U^2)  (\phi/\phi_0)$ where $\phi_0 = hc/e = 2.07\times 10^{-15}$Wb is the flux quantum, and $\phi=BA_0$ is the magnetic flux through a triangular plaquete of area $A_0 \approx (3.8\AA)^2 /2$ for the cuprate. At $B=10$T we find $\phi / \phi_0 \approx 3.5 \times 10^{-4}$. Further using commonly accepted values of $t_2=-0.3t$ , $U = 8t$ and $J=4t^2/U$,  we find $J_\chi \approx 5.6 \times 10^{-4} J$ at $B=10$T. The use of a smaller effective $U$ may increase this number a bit, but in any case a very small number is expected for $J_\chi$, due to the small ratio $\phi/\phi_0$. As we emphasized in this paper, the unexpected nature of the experimental data means that all avenues should be explored. Nevertheless, the small value of this term should be kept in mind. The assumed proximity to a quantum critical point also makes it challenging to explain the linear $B$ dependence of $\kappa_{xy}$ observed over a large range from 5T to 15T.

\acknowledgments

J. H. H. was supported by Samsung Science and Technology Foundation under Project Number SSTF-BA1701-07.  P. A. L. acknowledges support by DOE grant number DE-FG02-03ER46076.

\end{document}